\documentclass[11pt]{article}
\usepackage{moriond,epsfig}
\usepackage{psfig}
\usepackage{floatflt}

\def\Journal#1#2#3#4{{#1} {\bf #2}, #3 (#4)}

\def\NPB{{\em Nucl. Phys.} B}
\def\NPA{{\em Nucl. Phys.} A}
\def\PLB{{\em Phys. Lett.}  B}
\def\PRL{\em Phys. Rev. Lett.}
\def\PRC{{\em Phys. Rev.} C}

\def\JPG{{\em J. Phys.} G}

\def\be{\begin{equation}}
\def\ee{\end{equation}}
\def\bea{\begin{eqnarray}}
\def\eea{\end{eqnarray}}

\begin{document}
%\vspace*{4cm}
\title{Soft Particle Spectra at STAR}

\author{ Zhangbu Xu for the STAR Collaboration }

\address{Department of Physics, Building 510a, P.O. Box 5000,\\
Upton, NY 11973}

\maketitle\abstracts{
We presented the multiplicity and the spectra of many particles 
in Au+Au at $\sqrt{s_{_{NN}}}=130$ GeV measured by STAR detector.  
Their connections to initial condition, baryon creation, freeze-out 
condition and strangeness enhancement were discussed. 
}
{
  In relativistic heavy ion collisions at RHIC, thousands of particles 
of different properties are created. Although interactions with very 
high momentum transfer can be more easily calculated using perturbative 
QCD, the dominant interactions in these collisions happen at low momentum 
transfer. The bulk of the system may go through a phase transition from 
ordinary hadronic matter to hot and dense Quark-Gluon Plasma (QGP). 
To address this issue about whether a phase transition to QGP happens from 
measurements of soft particle spectra, we would like to know what the initial 
conditions are, whether the system is thermalized, whether the gluon degree 
of freedom dominates at certain stage, how long it takes for the system to 
evolve from one stage to another stage. 

We report the charged particle multiplicity, 
the mean transverse momentum ($p_{T}$) and its centrality dependence, the 
particle transverse momentum spectra of many particles  
($\pi^{\pm},K^{\pm}$, $K^{0}_{S},K^{\star},\phi,p,
\bar{p},\Lambda,\bar{\Lambda},\Xi,\bar{\Xi},\bar{d},\overline{^{3}He}$, 
{\it{etc.}}) and their centrality dependence measured by STAR detector at RHIC. 
Based on these measurements, we discussed the mean $p_{T}$ scaling in the context 
of gluon saturation model, the strangeness enhancement
(${\bar{\Lambda}}/{\bar{p}}$,$K/\pi$,$\phi/K^{\star}(892)^{0}$), 
radial flow (mass dependence of $m_{T}$ slope), the fitted results of statistical model 
and coalescence model on baryon chemical potential ($\mu_{B}$) and temperature 
($T$) and the time scale of the evolution of the system ($K^{\star}(892)^{0}/K$). 

The detector system used for these studies was the Solenoidal Tracker at RHIC 
(STAR). The main tracking device within STAR is the Time Projection Chamber 
(TPC) which is used to provide momentum information and particle 
identification for charged particles by measuring their ionization energy loss 
($dE/dx$). A minimum bias trigger was defined using coincidences between two 
Zero Degree Calorimeters (ZDC) which measured the spectator neutrons.  A 
Central Trigger Barrel (CTB) constructed of scintillator paddles surrounding 
the TPC was used to select small impact parameter ``central" collisions by 
selecting events with high charged particle multiplcity. Data were taken for 
Au+Au collisions at $\sqrt{s_{_{NN}}}=130$ GeV in the summer of 2000. 

STAR measured charge multiplicity per unit of pseudorapidity ($\eta$) for 
the 5\% most central collisions~\cite{hminus}. 
The negatively charged hadron multiplicity and its mean 
$p_{T}$ are $dN/d\eta|_{|\eta|<0.5}=567\pm1\pm38$ and 
$<p_{T}>=0.508\pm0.012$ GeV/$c$. The multiplicity per participant pairs 
increases by 38\% relative to $p\bar{p}$ and 52\% compared to nuclear 
collisions at $\sqrt{s_{_{NN}}}=17$ GeV. The $<p_{T}>$ increases by 30\% 
and 18\% respectively from $0.392$ GeV/$c$ in $p\bar{p}$ collisions and 
$\simeq0.429$ GeV/$c$ in Pb+Pb collisions at SPS~\cite{hminus,na49}. 
\begin{figure}
\begin{minipage}{3.7in}
  {\includegraphics*[width=1.8in]{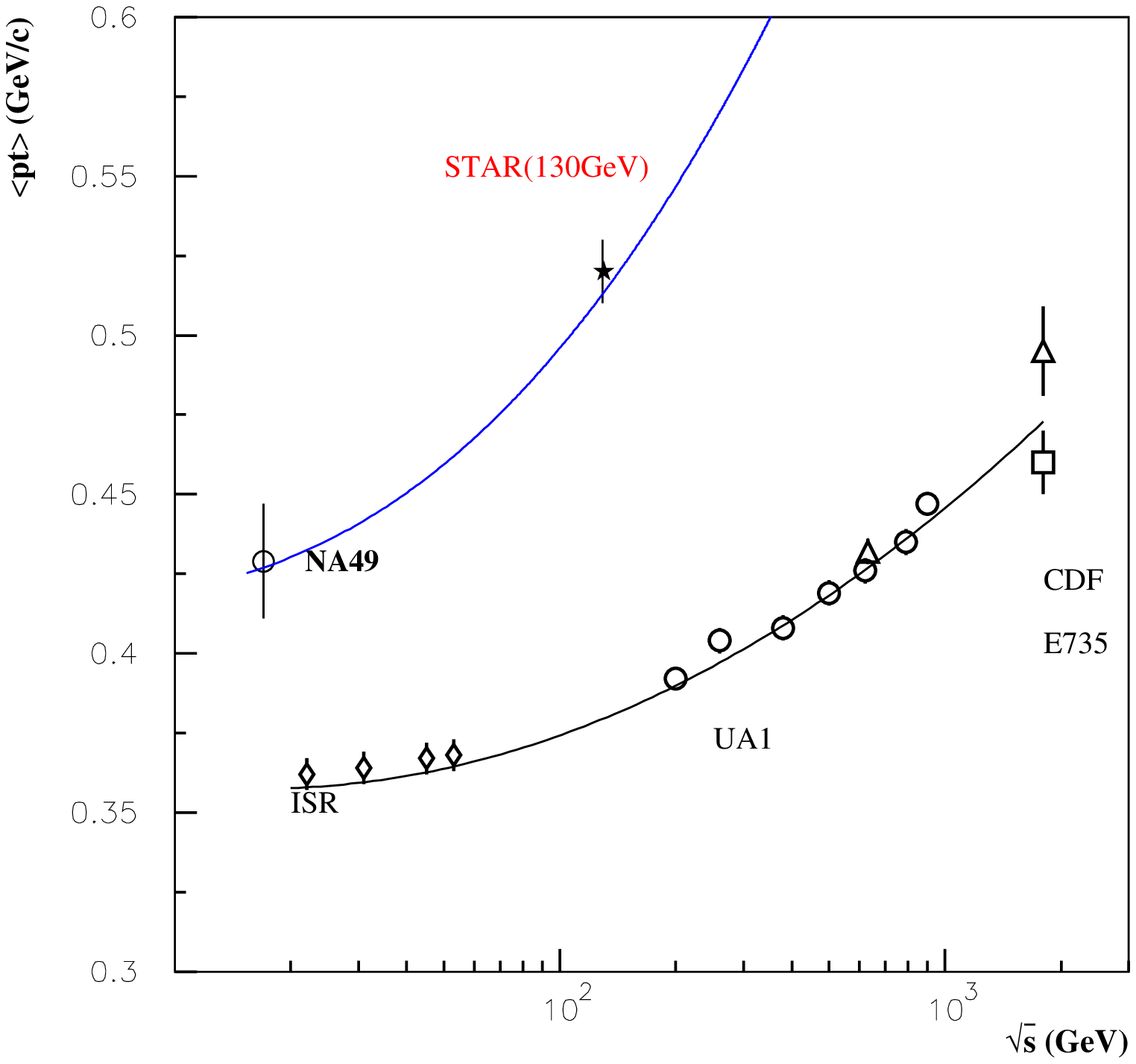}} 
  {\includegraphics*[width=1.8in]{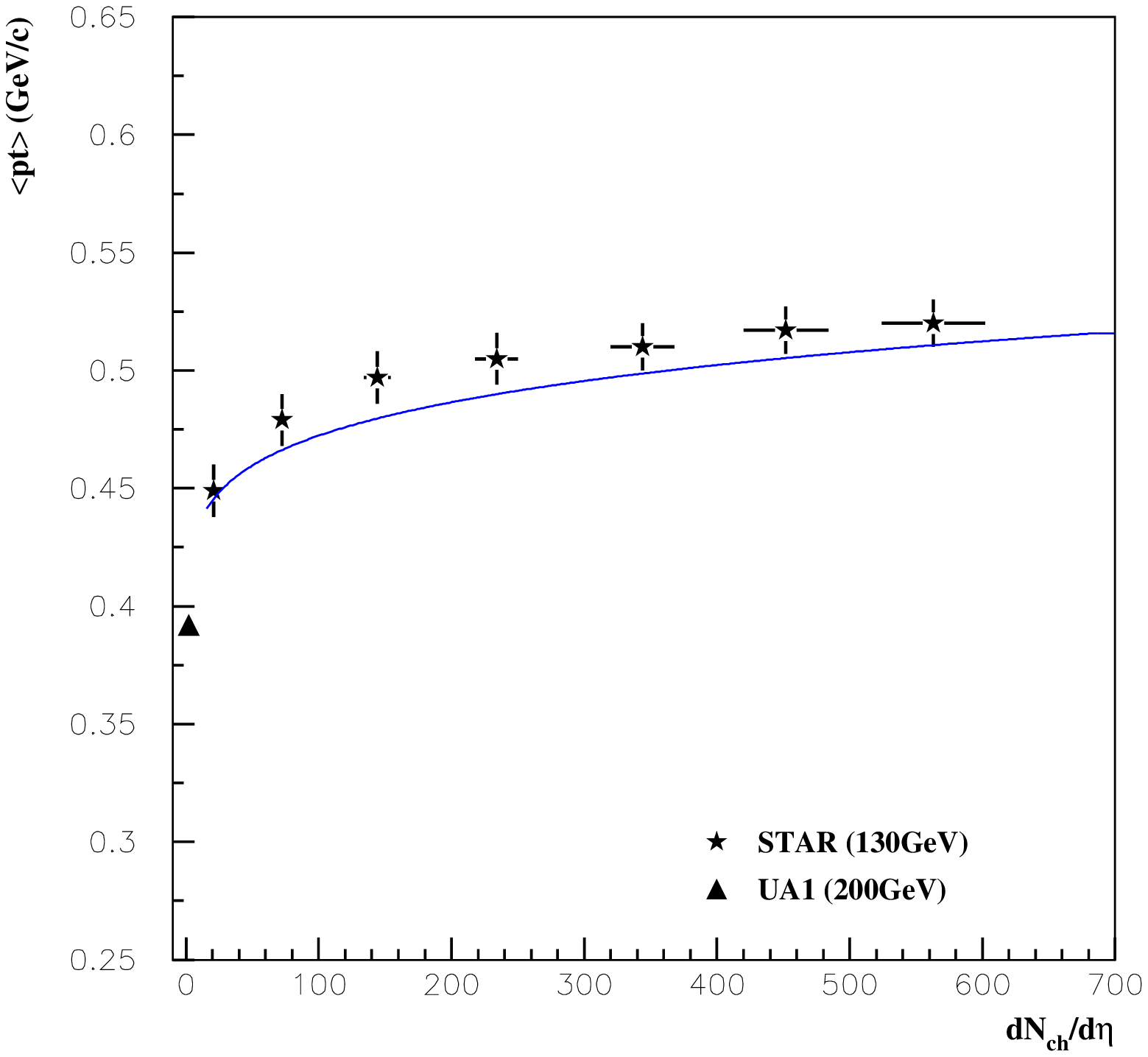}} 
  \caption{Left: mean {\protect{$p_{T}$}} of \protect{$h^{-},N_{ch}$} as a 
  function of \protect{$\sqrt{s_{_{NN}}}$} for \protect{$pp,\bar{p}p$ and 
  central $AA$ collisions}.
  Right: mean {\protect{$p_{T}$}} of \protect{$N_{ch}$} as a 
  function of \protect{$N_{ch}$} at \protect{$\sqrt{s_{_{NN}}}=130$ GeV.} 
  Curves are from Eq.1.} \label{ptscaling} 
\end{minipage}
\begin{minipage}{2.0in}
  {\includegraphics*[width=1.8in]{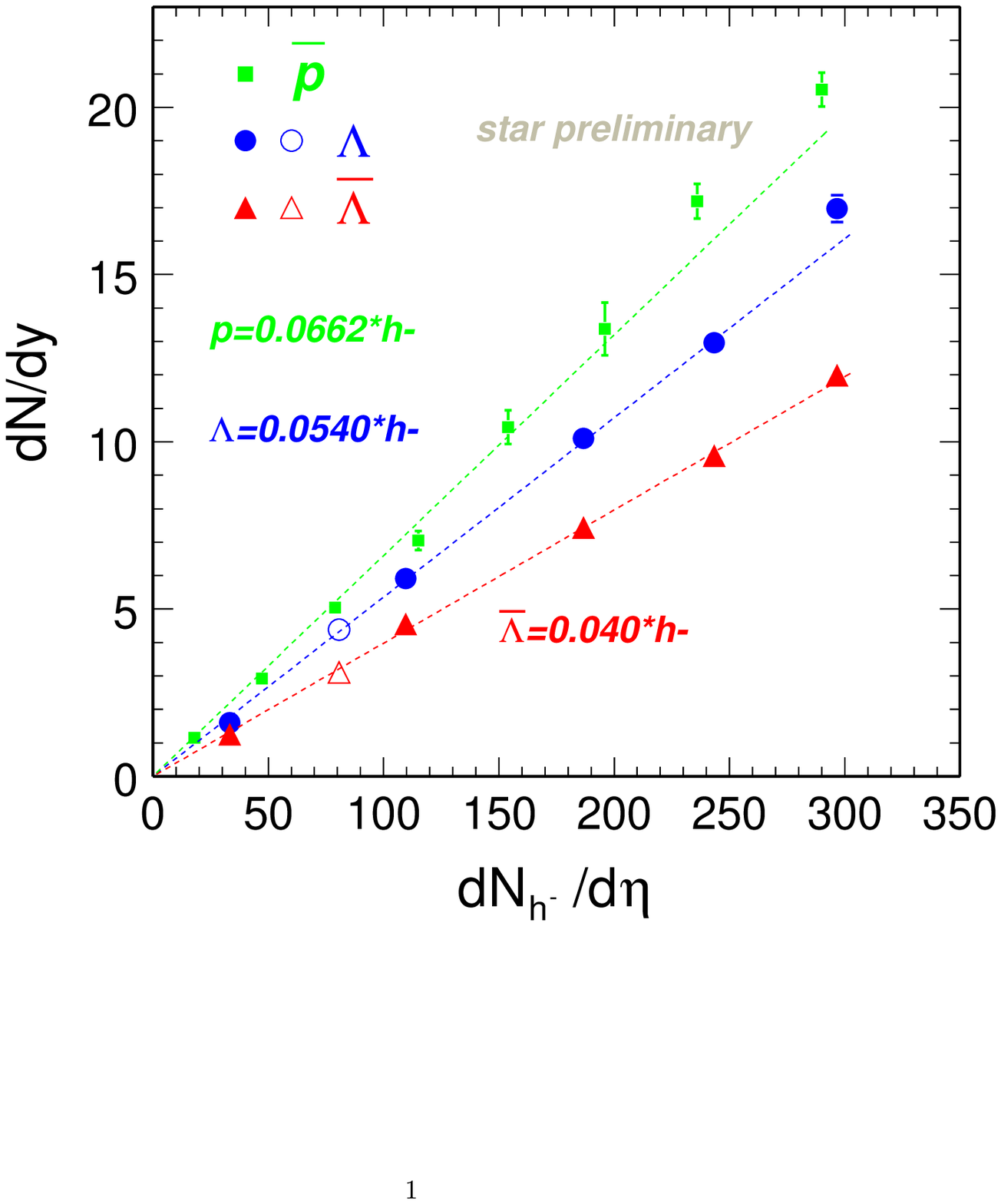}} 
\caption{$dN/dy$ of $\bar{p}$ ,$\Lambda$, $\bar{\Lambda}$ as a function 
of $h^{-}$. Linear dependence can describe the data well. 
} \label{lambda} 
\end{minipage}
\end{figure} 
The left panel of Fig.~\ref{ptscaling} shows mean {$p_{T}$} of $h^{-}$ or 
$N_{ch}$ as a function of $\sqrt{s_{_{NN}}}$ for $pp,\bar{p}p,AA$ collisions. 
$<p_{T}>_{pp} = 0.40-0.03\times\ln{x}+0.0053\times(\ln{x})^{2}$ is a 
parametrization of the measured mean $p_{T}$ in pp collisions~\cite{ua1} 
as shown in the plot. Both gluon saturation~\cite{schaffner1} and 
hydrodynomics~\cite{hydroscaling} predict some scaling behavior of 
$<p_{T}>\sim \sqrt{({{dN}\over{d\eta}})_{_{AA}}/{\pi{R^{2}}}}$ when conditions 
are satisfied. To get a quantitative analysis on the dependence between 
$<p_{T}>$ and multiplicity, we require that the scaling dependence is 
linear and it should also satisfy the $pp$ results. By doing this, 
there is only one free parameter as shown in Eq.~\ref{eq:saturation}. 
\begin{equation}
<p_{T}>_{_{AA}} = a+\sqrt{{s_{_{AA}}}\over{s_{pp}}}(<p_{T}>_{pp}-a),
\label{eq:saturation}
\end{equation}
where 
$s_{_{AA}} = {({{dN}\over{d\eta}})_{_{AA}}/{\pi{R^{2}}}}$ and 
$s_{pp} = {({{dN}\over{d\eta}})_{pp}/{\pi{r_{0}^{2}}}}$ are the multiplicity 
density per unit pseudorapidity per unit tranverse area in 
AA and pp collisions($R=r_{0}A^{1/3}$) and 
$a=0.3$ GeV/$c$ is a constant chosen to describe the AA data. 
For non-central collisions, calculation of the transverse area is not 
as straight forward. We take the parametrization of the centrality 
dependence of $s_{_{AA}}$ from Ref.~\cite{schaffner1}. The solid curves 
in Fig.~\ref{ptscaling} show the energy dependence and centrality 
dependence of $<p_{T}>$ from Eq.~\ref{eq:saturation}. The curves describe 
the data well. It will be interesting to see if the description holds with  
an energy scan from lower energy to highest energy at RHIC. 

Strangeness enhancement has long been predicted to be a possible signature of 
QGP~\cite{rafelski1}. The measurements of kaon, $\Lambda$, $\phi$, $\Xi$, 
$\Omega$ and their antiparticles have been under way with the STAR detector. 
Measurements of antihyperon to antiproton ratio 
($\bar{\Lambda}/\bar{p}$) at lower energies indicate large enhancement 
of this ratio in AA collisions when compared to $pp$ collisions~\cite{e917}. 
The interpretation of the results is complicated by the fact that the antimatter 
absorption in a baryon rich environment is significant or even dominant at lower 
beam energies~\cite{wang}. At RHIC, the $\bar{p}/p$ ratio is approaching unit indicating 
low net baryon density~\cite{pbar}. Both $\bar{p}$ and $\bar{\Lambda}$ were 
measured at STAR with high statistics~\cite{pbar,lambdabar}  
as shown in Fig.~\ref{lambda}. 
Antiprotons which include feeddown from $\bar{\Lambda}$ decays are identified by the 
ionization energy loss ($dE/dx$) when particles traverse the TPC. 
Antilambdas are identified by their decay topology and mixed-event method. The 
integrated yields of $dN/dy$ at $y=0$ of the three particles are approximately 
proportional to the $dN_{h^{-}}/d\eta$. This means that there is very little 
centrality dependence of $\bar{p}/\pi$ and $\bar{\Lambda}/\bar{p}$ which is very 
different from AGS and SPS results~\cite{e917,na44}. Our result of 
$\bar{\Lambda}/\bar{p}=0.93^{+0.57}_{-0.23}$ after feeddown 
correction~\cite{pbar,lambdabar} for both $\bar{\Lambda}$ and $\bar{p}$ in central 
Au+Au collisions shows significant increase from elementary collisions. 
It has been speculated that the enhancement seen in pA and AA collisions may be related 
to number of collisions~\cite{e910} or the degree of violence of the collisions 
(characterized by charge multiplicity)~\cite{ua1strange}. An alternative explanation 
is that strangeness is in equilibrium with the rest of the particles in 
the system since the thermal/statistical models describe the data well at 
RHIC~\cite{magestro,nxu}. In addition, $K/\pi$ and $\phi/K^{\star0}$ measured at 
STAR~\cite{starstrange} also indicate enhancement in AA collisions relative to 
elementary collisions. 

The detailed $p_{T}$ spectra of identified particles carry information about the radial 
flow and temperature of the system~\cite{nxu,teaney}. 
Fig.~\ref{mass_flow} (left panel) shows 
$p_{T}$ spectra of pseudoscalar and vector meson, hyperon and strange meson, baryon and 
meson. They exhibit a common feature that the production of heavy particle decreases 
much slower than that of lighter particle with increasing $p_{T}$. At $p_{T}\simeq2$ 
GeV/c, there are as many heavy particles produced as the lighter particles. This can be 
explained by the radial flow where all particles have same temperature and a common 
flow velocity~\cite{nxu}. Fig.~\ref{mass_flow} (right panel) shows the inverse slope 
parameters of $p_{T}$ spectra as a function of particle mass. It has been 
seen that heavier particles have larger flow effect. 
In fact, thermal model with radial flow can extract the 
temperature and flow velocity at kinetic freeze-out from the $p_{T}$ spectra at RHIC. 
There is also a novel mechanism proposed to explain the $p/\pi$ ratio~\cite{gyulassy}.
\begin{figure}
\begin{minipage}{5.5in}
  {\includegraphics*[width=3.in]{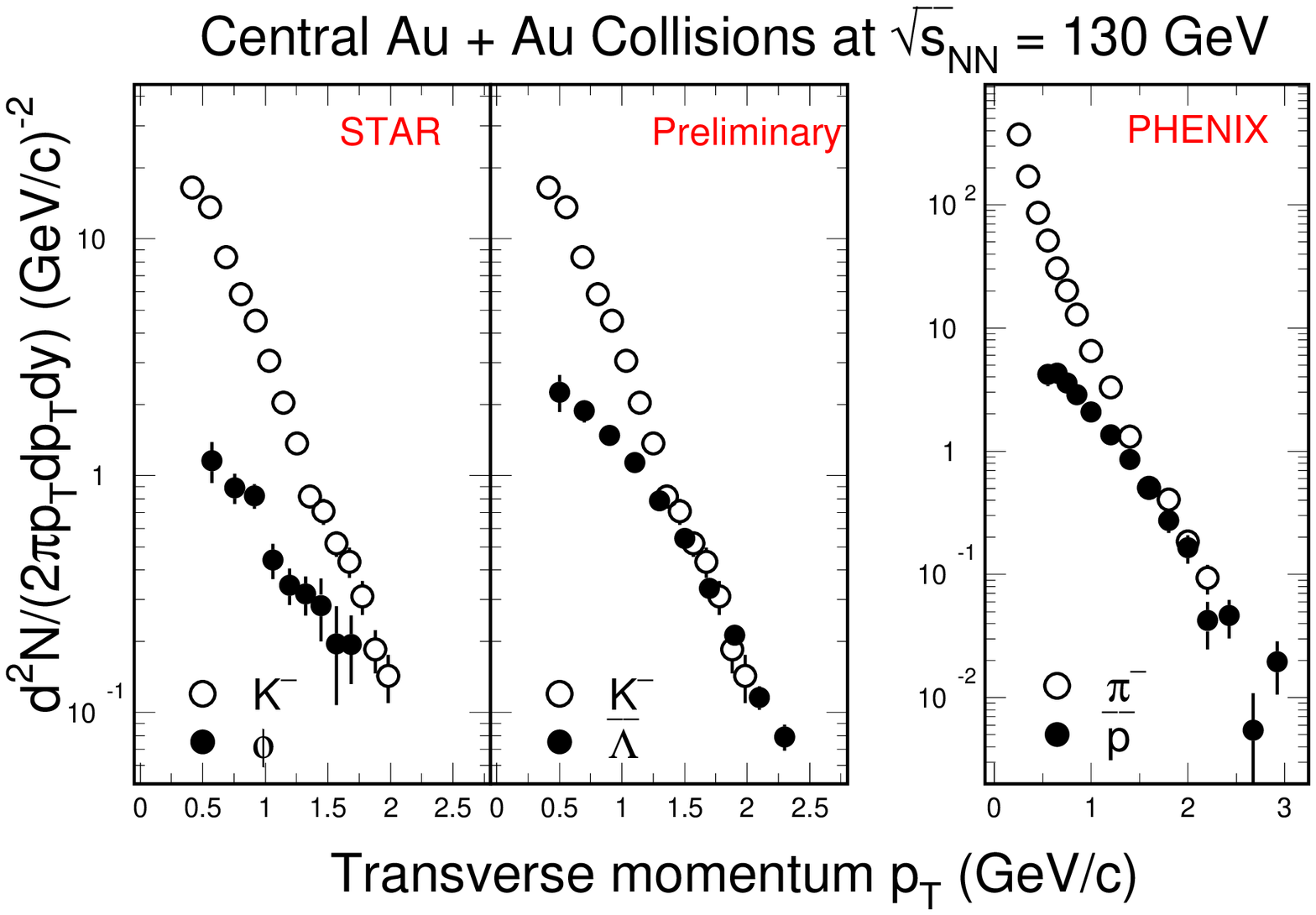}}
  {\includegraphics*[width=2.5in]{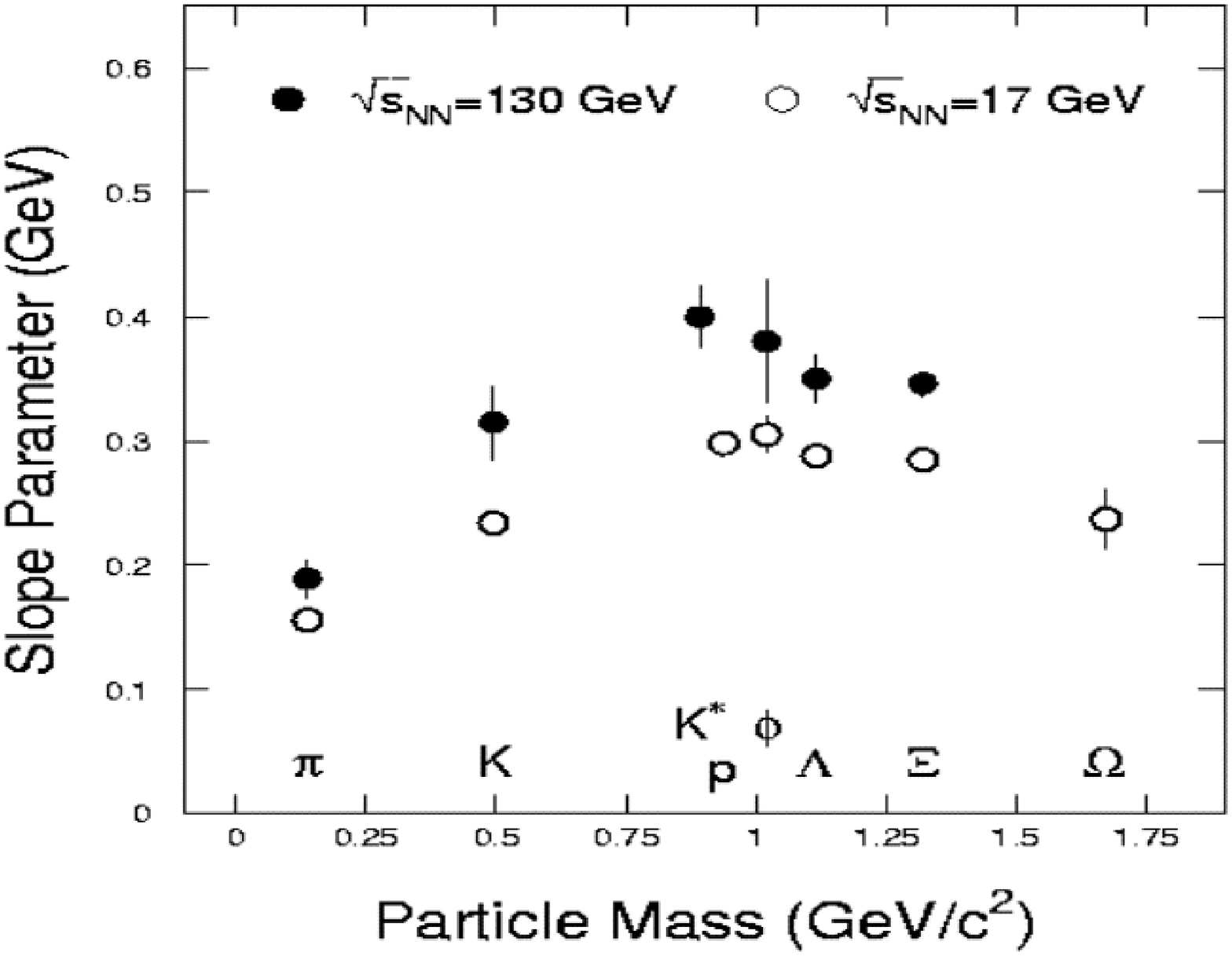}}
  \caption{Left: $p_{T}$ spectra of pseudoscalar and vector meson, 
  hyperon and strange meson, baryon and meson. 
  Right: Inverse slope of the transerve spectra vs mass of the particle. The errors are 
  statistical only. The systematical uncertainty of STAR data is 10-15\% depending on 
  particles.} 
  \label{mass_flow} 
\end{minipage}%
%\vspace{-0.7cm}
\end{figure} 

One of the unknowns of particle production in strong interaction is the 
probability that a quark and antiquark pair produces a vector meson as compared 
to its pseudoscalar partner. It is thus important to measure the vector mesons and 
pseudoscalar mesons to understand the production mechanism involving spin in strong 
interactions under extreme conditions. We have measured the $K^{\star}(892)^{0}$ and 
$\overline{K}^{\star}(892)^{0}$ in relativistic nucleus-nucleus collisions~\cite{kstar} as shown in Fig.~\ref{kstar}. 
The $K^{\star0}$ $m_{T}$ spectrum from the 14\% most central Au+Au collisions results in 
an inverse slope parameter of $0.40\pm0.02\pm0.04$ GeV, similar to that measured for the 
$\phi$ meson using a similar centrality cut~\cite{starstrange}. The measured yield, $dN/dy=10.0\pm0.9$(stat) 
$\pm2.5$(sys) is relatively high compared to elementary collisions and thermal model 
predictions, considering the short $K^{\star0}$ lifetime ($c\tau\simeq4$ fm) and expected 
losses due to rescattering of the decay daughters in the dense medium~\cite{kstar}. 
\begin{figure}
\begin{minipage}{4.in}
  {\includegraphics*[width=1.9in]{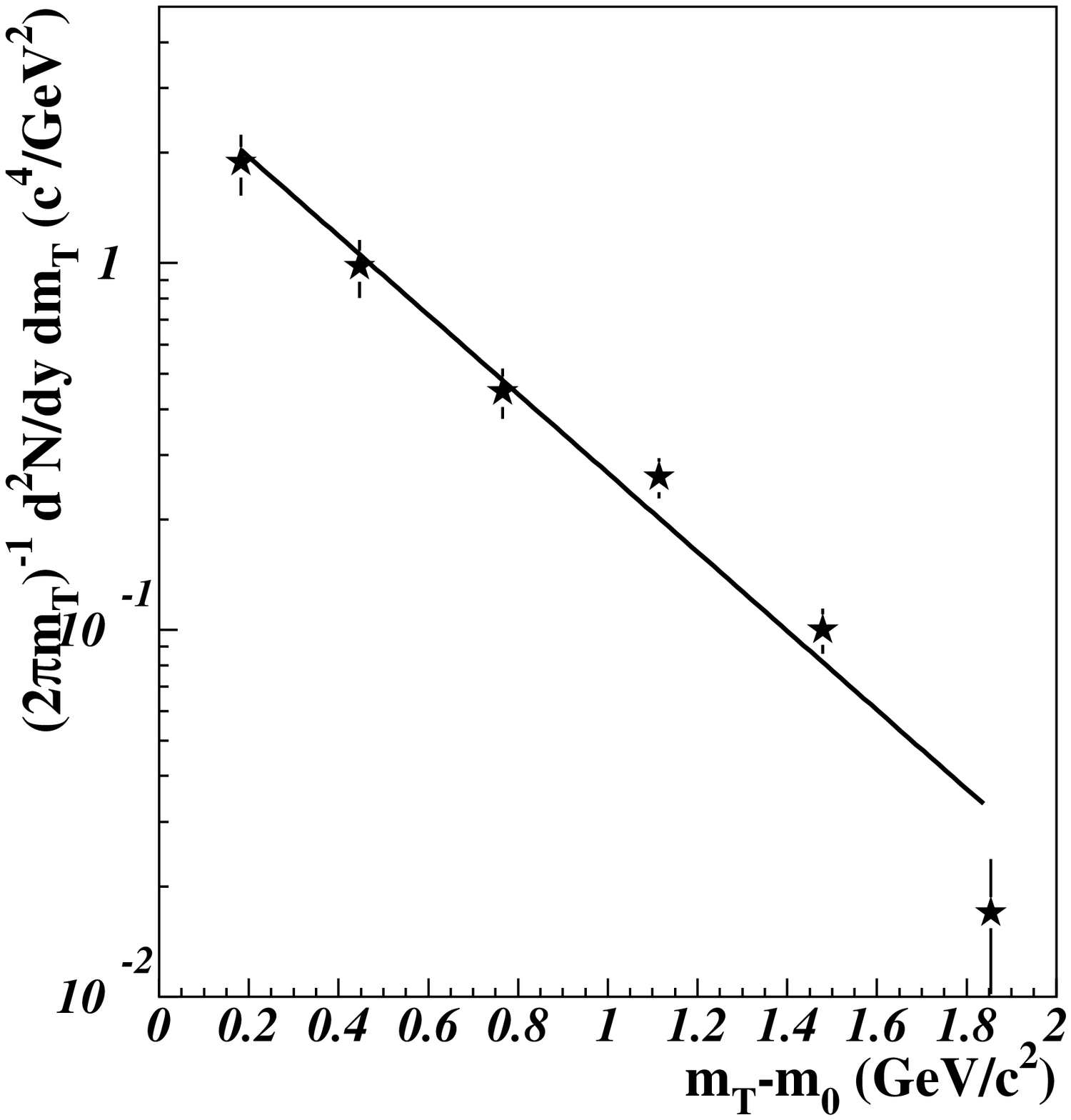}} 
  {\includegraphics*[width=2.1in]{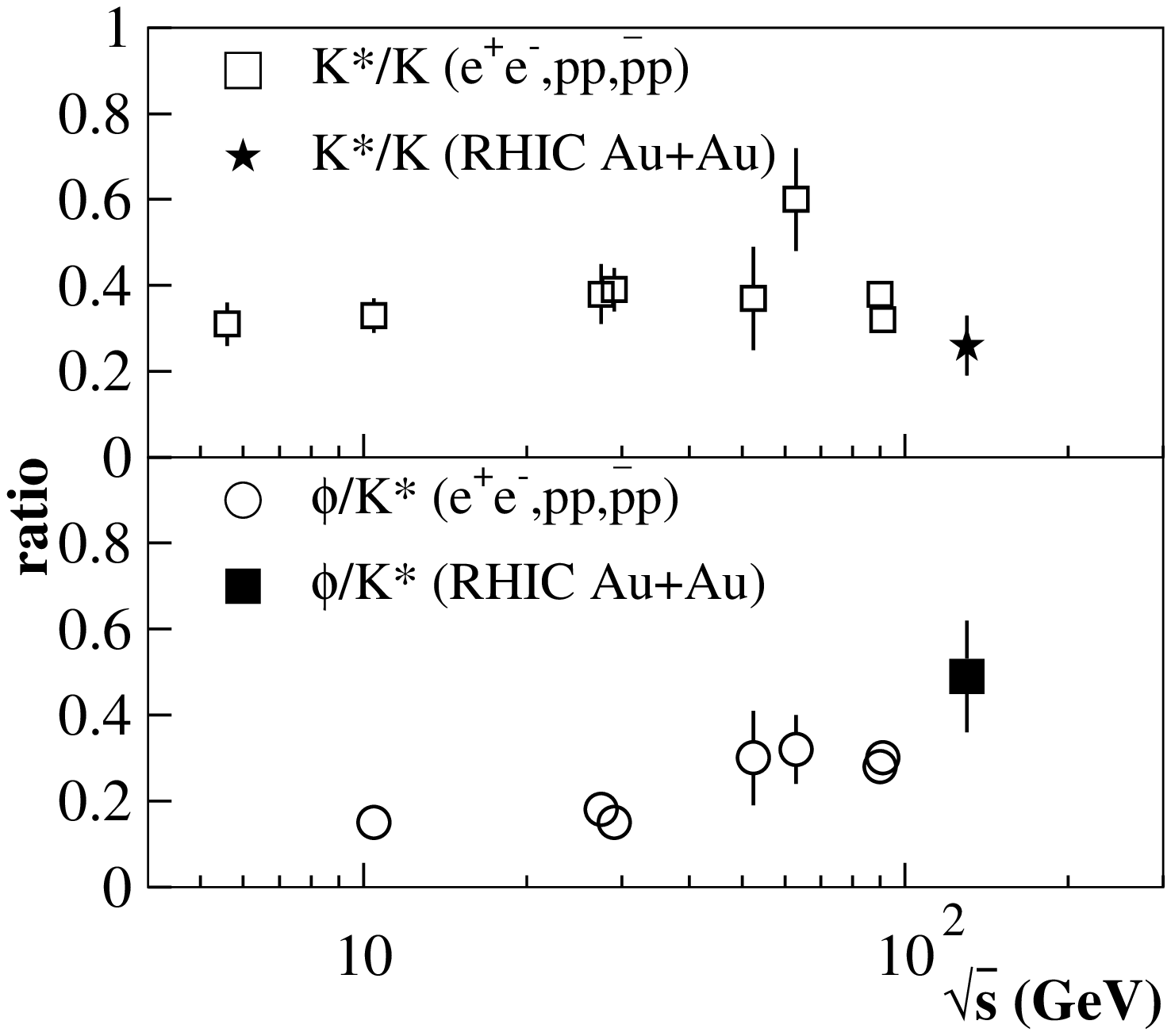}} 
  \caption{Left: $m_{T}$ spectrum of $K^{\star0}$ in 14\% most central Au+Au collisions. 
  Right: $K^{\star0}/K$ and $\phi/K^{\star0}$ ratios vs beam energy in elementary 
collisions and central Au+Au collisions. } \label{kstar} 
\end{minipage}
\begin{minipage}{2.in}
  {\includegraphics*[width=2.in]{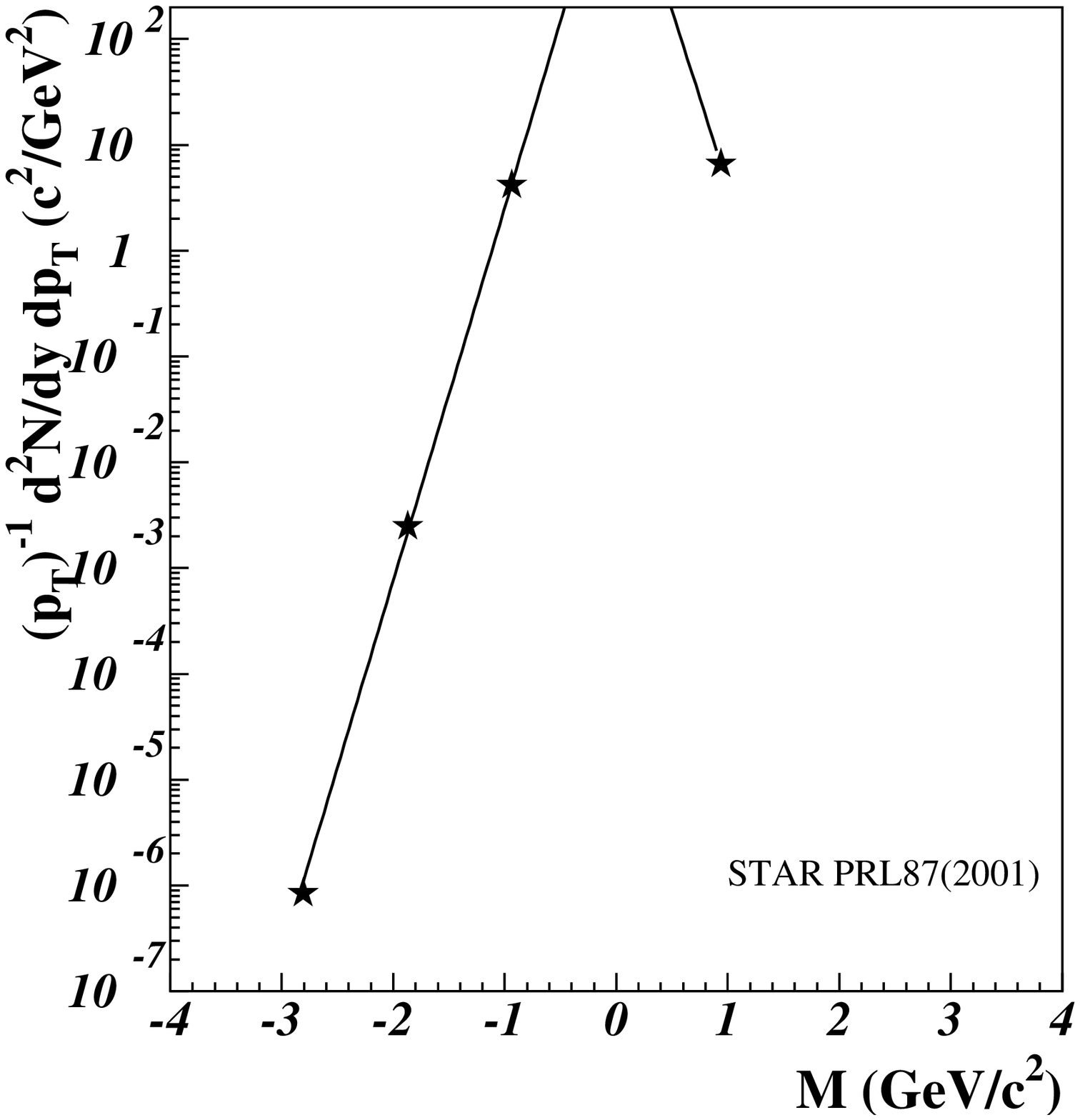}} 
  \caption{The measured differiential yield at $p_{T}\simeq0.4$ GeV/$c$ as 
function of particle's atomic number $A$ for $p,\bar{p},\bar{d},\overline{^{3}He}$. 
} \label{yield_vs_a} 
\end{minipage}%
\end{figure} 

The measurements of nuclei and antinuclei production are sensitive to the volume and 
nucleon and antinuleon phase space density at the freeze-out because the nuclei are 
believed to be produced via coalescence of nucleons~\cite{stardbar}. Due to secondary 
nuclei production from beam pipe and limited $p_{T}$ range of PID from $dE/dx$, STAR is 
only able to identify and measure light antinuclei ($\bar{d},\overline{^{3}He}$), 
$\bar{p}$ and proton. Fig.~\ref{yield_vs_a} shows the measured differiential 
yield at $p_{T}/A\simeq0.4$ GeV/$c$ as function of particle's atomic number $A$ for 
$p,\bar{p},\bar{d},\overline{^{3}He}$. Using a simplified formulae 
${{dN}\over{p_{T}dp_{T}dy}}{\propto}\exp{(-(m_{B}\pm\mu_{B})A/T)}$ from 
Ref.~\cite{heinzcoalescence}, we are able to fit well the nucleus production at low 
$p_{T}$ at AGS~\cite{e864xzb}, SPS~\cite{na52} and RHIC heavy ion collisions and obtain temperature and baryon 
chemical potential at kinetic freeze-out. The results are: $T=126MeV$, $\mu_{B}=21MeV$ 
(RHIC); $T=130MeV$, $\mu_{B}=170MeV$ (SPS); $T=110MeV$, $\mu_{B}=500MeV$ (AGS). Errors 
were estimated to be $\pm10MeV$ for all the cases~\cite{e864xzb}. The ratios of $\mu_{B}/T$ at kinetic freeze-out vs that at chemical 
freeze-out at AGS, SPS and RHIC central AA collisions~\cite{magestro} are  
consistent with each other even though the parameters at chemical freeze-out are 
extracted from thermal fit to the many ratios of the integrated particle yields while 
the parameters from coalescence are from nuclear cluster formation rate at low $p_{T}$. 
However, further study is needed to see if such a trajectory in the phase diagram is 
possible~\cite{rapp}.

In summary, we report recent results from the STAR experiment: i) Charged particle 
multiplicity and their averaged values of transverse momentum are discussed within the 
framework of both gluon saturation and hydrodynamic models; ii) The identified particle 
distributions provided information on both chemical equilibrium and collective expansion; 
iii) Resonances and light nuclei formation may provide a probe of the 
evolution of the system at late stage of the collision. In order to determine the formation of a system with partonic 
degree of freedom, we need to systematically measure the spectra of all the particles 
listed and new particles (especially multistrange and heavy-flavor particles) at 
different beam energies and beam species. 
}
\footnotesize
{

}
\end{document}